\documentstyle[seceq,preprint]{jpsj}

\def\runtitle{A DM Algorithm for 3D Classical Models}
\def\runauthor{Tomotoshi {\sc Nishino} and Kouichi {\sc Okunishi}}

\title{A Density Matrix Algorithm for 3D Classical Models}

\author
{ Tomotoshi {\sc Nishino}\footnote{e-mail: 
nishino@phys560.phys.kobe-u.ac.jp} and 
Kouichi {\sc Okunishi}$^{1,}$\footnote{e-mail: 
okunishi@godzilla.phys.sci.osaka-u.ac.jp} }

\inst
{ Department of Physics, Faculty of Science, 
Kobe University, Rokkodai 657-8501\\
$^1$Department of Physics, Graduate School of Science, 
Osaka University, Toyonaka 560-0043 }

\abst{
We generalize the corner transfer matrix renormalization group, which
consists of White's density matrix algorithm and Baxter's
method of the corner transfer matrix, to three dimensional (3D) classical
models. The renormalization group transformation is obtained 
through the diagonalization of density matrices for a cubic cluster.
A trial application to 3D Ising model with $m = 2$ is shown as the
simplest case.
}

\kword
{ Density Matrix, Renormalization Group, Corner Tensor. }

\begin{document}
\sloppy
\maketitle


\section{Introduction}

Variational estimation of the partition function has been one of the
standard technic in statistical mechanics. For a
two-dimensional (2D) classical lattice model defined by a transfer matrix
$T$, the variational partition function per row is written as
\begin{equation}
\lambda = \frac{\langle V | \, T \, | V \rangle}{\langle V | V \rangle} \, ,
\end{equation}
where $| V \rangle$ represents the trial state and $\langle V |$ is its
conjugate; $\lambda$ is maximized when $\langle V |$ and $| V \rangle$
coincide with the left and the right eigenvectors of $T$, respectively. In
1941 Klamers and Wannier\cite{Krm,Kik} investigated the Ising model,
assuming that $| V \rangle$ is well approximated by a product of
matrices  
\begin{equation}
V(\ldots, i, j, k, l,\ldots) 
= \ldots F^{ij}_{~} F^{jk}_{~} F^{kl}_{~} \ldots \, ,
\end{equation}
where $i, j, k, l$, etc., are the Ising variables, and $F^{ij}_{~}$ is a
symmetric 2 by 2 matrix. The approximation is more accurate than
both the mean-field and the Bethe
approximations.~\cite{Bethe} Baxter improved the trial state by
introducing additional degrees of freedom.~\cite{Bx1,Bx2,Bx3} His
variational state is defined as  
\begin{equation}
V(\ldots, i, j, k, l,\ldots) 
= \sum_{\ldots, a, b, c, d, \ldots} 
\ldots F^{ij}_{ab} F^{jk}_{bc} F^{kl}_{cd} \ldots \, , 
\end{equation}
where $a, b, c, d$, etc., are $2^n_{~}$-state group spin variables. The
tensor $F^{ij}_{ab}$ contains $4 \cdot 2^{2n}_{~}$ elements, and
therefore it is not easy to optimize $F^{ij}_{ab}$ --- adjust the elements
--- so that $\lambda$ is maximized. He solved the optimization problem
by introducing the corner transfer matrix (CTM), and by solving 
self-consistent equations for the tensors.~\cite{Bx3} In 1985
Nightingale and Bl\"ote used Baxter's tensor product as a initial state in
the projector Monte Carlo simulation for the Haldane system.~\cite{NB}
Baxter suggested an outline of generalizing his CTM method to 3D
systems,~\cite{Bx3} however, the project has not been completed.

Similar variational formulations have been applied to one-dimensional
(1D) quantum systems, especially for $S = 1$ spin chains. The variational
ground state  $| \Psi \rangle$ is given by a modified tensor product
\begin{equation}
\Psi (\ldots, i, j, k, l, \ldots) = \sum_{\ldots, a, b, c, d, e, \ldots}
\ldots A^i_{ab} A^j_{bc} A^k_{cd} A^l_{de} \ldots \, ,
\end{equation}
where the subscripts $a, b, c, d, e,$ etc., are $m$-state group spin
variables.  Affleck, Lieb, Kennedy, and Tasaki (AKLT) showed that the
ground-state of a bilinear-biquadratic $S = 1$ spin chain is exactly
expressed by the tensor product with $m = 2$.~\cite{AKLT} The
variational formulation has been generalized by Fannes et. al.
for the arbitrary large $m$.~\cite{Fannes1,Fannes2,Fannes3} Now such
ground state is called `finitely correlated state'~\cite{Fannes2} or
`matrix product state'.~\cite{Zitt1} Quite recently Niggemann et.
al.~\cite{Zitt2} showed that the ground state of a 2D quantum systems
can be exactly written in terms of a two-dimensional tensor product.
Although $| \Psi \rangle$ in Eq.(1.3) does not look like $| V \rangle$ in
Eq.(1.4), they are essentially the same. We can transform $| V \rangle$
into $| \Psi \rangle$ by obtaining $A^i_{ab}$ from $F^{ij}_{ab}$ through a
kind of duality transformation;~\cite{Takas} the opposite is also possible.

The application of both $| V \rangle$ in Eq.(1.3) and $| \Psi \rangle$ in
Eq.(1.4) are limited to translationally invariant (or homogeneous)
systems. In 1992 White established a more flexible numerical variational
method from the view point of the real-space renormalization group
(RG).~\cite{Wh1,Wh2} Since his numerical algorithm is written in terms
of the density matrix, the algorithm is called `density matrix
renormalization group' (DMRG). White's variational state is written in a
position dependent tensor product~\cite{Ostlund1,Ostlund2}
\begin{equation}
\Phi (\ldots, i, j, k, l, \ldots) = \sum_{\ldots, a, b, c, d, e, \ldots}
\ldots A^i_{ab} B^j_{bc} C^k_{cd} D^l_{de} \ldots \, ,
\end{equation}
where $A^i_{ab}$ is not always equal to $B^i_{ab}$, etc. This
inhomogeneous property in $| \Phi \rangle$ makes DMRG possible to treat
open boundary systems~\cite{Huse} and random systems.~\cite{Hida} Now
the DMRG is widely used for both quantum~\cite{Escorial} and 
classical~\cite{Ni,Carlon1,Carlon2} problems. Quite recently Dukelsky et.
al. analyzed the correspondence (and a small discrepancy) between DMRG
and the variational formula in
Eq.(1.4).~\cite{RecGerman0,RecGerman1,RecGerman2}

The decomposition of the trial state into the tensor product tells us how
to treat lattice models when we try to obtain the partition function. The
essential point is to break-up  the system into smaller pieces --- like the
local tensor $F^{ij}_{ab}$ in Eq.(1.3) or $A^i_{ab}$ in Eq.(1.4) --- and
reconstruct the original system by multiplying them. According to
this idea, the authors combine DMRG and Baxter's method of CTM, and
proposed the corner transfer matrix renromalization group (CTMRG)
method.~\cite{CTMRG1,CTMRG2} It has been shown that CTMRG is
efficient for determinations of critical indices~\cite{CTMRG2} or latent
heats.~\cite{q5} 

The purpose of this paper is to generalize the algorithm of CTMRG to
three-dimensional (3D) classical systems. We focus on the RG algorithm
rather than its practical use. In the next section, we briefly review the
outline of CTMRG. The key point is that the RG transformation is obtained
through the diagonalization of the density matrix. In \S 3 we define the
density matrix for a 3D vertex model, and in \S 4 we explain the way to
obtain the RG transformation. The numerical algorithm is shown in \S 5.
A trial application with $m = 2$ is shown for the 3D Ising Model.
Conclusions are summarized in \S 6.

\section{Formulation in Two Dimension}

\begin{figure}
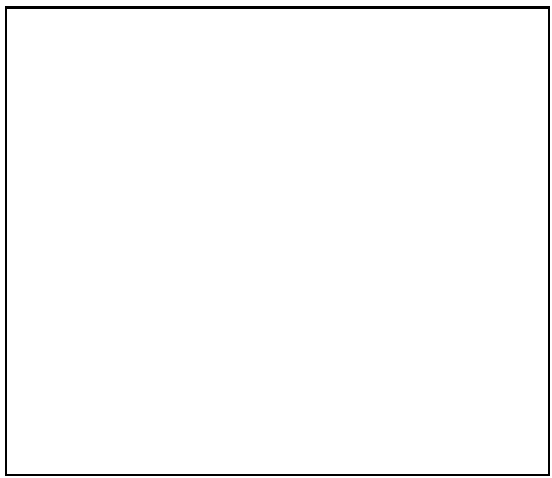

\figureheight{6cm}
\caption{
Square cluster of a symmetric vertex model; the shown system is the
example with linear dimension $2N = 6$. The cross marks $\times$ show
the boundary spins.
 } 
\label{fig:1}
\end{figure}

The aim of CTMRG is to obtain variational partition
functions of 2D classical models. Let us consider a square cluster of
a 16-vertex model (Fig.1) as an example of 2D systems. We
impose the fixed boundary condition, where the boundary spins shown by
the cross marks point the same direction. In order to
simplify the following discussion, we assign a symmetric Boltzmann
weight $W^{~}_{ijkl} = W^{~}_{jkli} = W^{~}_{ilkj}$ to each
vertex,~\cite{Simple} where $i,j,k$ and $l$ denote  two-state spins (=
Ising spins or arrows) on the bonds. (Fig.2(a)) 

\begin{figure}
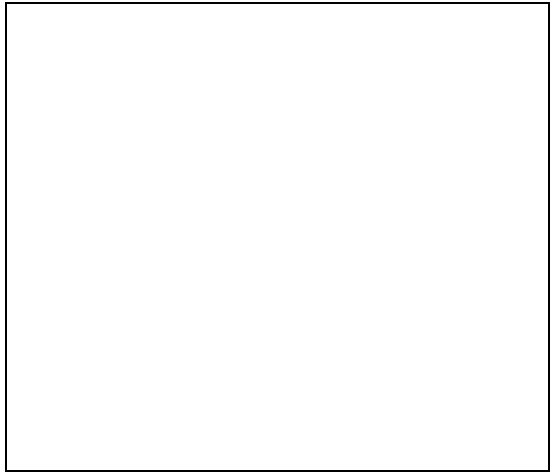

\figureheight{6cm}
\caption{
Boltzmann weight and transfer matrices. The dots represent
spin variables inside the square cluster shown in Fig.1, and the cross
marks represent the boundary spins. 
(a) Vertex weight $W^{~}_{ijkl}$. 
(b) Half-row transfer matrix $P^{~i}_{ab}$.
(c) Corner transfer matrix $C^{~}_{ab}$.
 } 
\label{fig:2}
\end{figure}

We employ two kinds of transfer matrices in order to express the
partition function $Z_{2N}^{~}$ of the square cluster with linear
dimension $2N$.  One is the half-row transfer matrix (HRTM). Figure 2(b)
shows the HRTM $P^{~i}_{ab}$ with length $N = 3$, where the subscripts
$a = (a_1^{~}, a_2^{~}, \ldots, a_N^{~})$ and $b = (b_1^{~}, b_2^{~}, \ldots,
b_N^{~})$, respectively, represent row spins --- in-line spins --- on the
left and the right sides of the HRTM. We think of $P^{~i}_{ab}$ as a matrix
labeled by the superscript $i$. The other is the Baxter's
corner transfer matrix (CTM),~\cite{Bx1,Bx2,Bx3} that represents 
Boltzmann weight of a quadrant of the square. Figure 2(c) shows the CTM
$C^{~}_{ab}$ with linear dimension $N = 3$. The partition function
$Z_{2N}^{~}$ is then expressed as
\begin{equation}
Z_{2N}^{~} = {\rm Tr} \, \rho = {\rm Tr} \, C^4_{~} \, ,
\end{equation}
where $\rho^{~}_{ab} \equiv \left( C^4_{~} \right)^{~}_{ab}$ is the density
matrix. From the symmetry of the vertex weight $W^{~}_{ijkl}$, the
matrices $P^{~i}_{ab}$, $C^{~}_{ab}$, and $\rho^{~}_{ab}$ are invariant
under the permutation of subscripts.

\begin{figure}
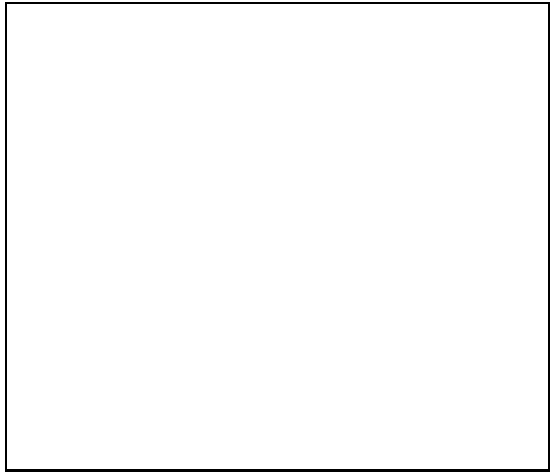

\figureheight{6cm}
\caption{
Extensions of (a) the HRTM (Eq.(2.2)), and (b) the CTM. (Eq.(2.3))
 } 
\label{fig:3}
\end{figure}

There are recursive relations between $W$, $P$, and $C$. 
We can increase the length of HRTM by joining a vertex
\begin{equation}
{P}^{~i}_{{\bar a}{\bar b}} = 
\sum_k W^{~}_{ijkl} P^{~k}_{ab} \, ,
\end{equation}
where the extended row-spins are defined as 
${\bar a} = (a, l) = (a_1^{~}, a_2^{~}, \ldots, a_N^{~}, l)$ and 
${\bar b} = (b, j) = (b_1^{~}, b_2^{~}, \ldots, b_N^{~}, j)$. (Fig.3(a))
In the same manner, the area of CTM can be extended by joining two HRTMs
and a vertex to the CTM
\begin{equation}
C^{~}_{{\bar a}{\bar b}} = 
\sum_{cd~kj} W^{~}_{ijkl} P^{~j}_{db} P^{~k}_{ac} C^{~}_{cd} \, ,
\end{equation}
where the extended row-spins ${\bar a}$ and ${\bar b}$ are defined as 
${\bar a} = (a, l) = (a_1^{~}, a_2^{~}, \ldots, a_N^{~}, l)$ and 
${\bar b} = (b, i) = (b_1^{~}, b_2^{~}, \ldots, a_N^{~}, i)$. (Fig.3(b))
In this way, we can construct HRTM and CTM with arbitrary size $N$ by
repeating the extension Eqs.(2.2) and (2.3).

It should be noted that the matrix dimension of both $C^{~}_{ab}$ and
$P^{~i}_{ab}$ increases very rapidly with their linear size $N$. The fact
prevents us to store the matrix elements of $C^{~}_{ab}$ and $P^{~i}_{ab}$
when we numerically calculate the partition function $Z_{2N}^{~}$.
This difficulty can be overcomed by compressing CTM and HRTM into
smaller matrices via the density matrix algorithm,~\cite{Wh1,Wh2} where
the RG transformation is obtained through the diagonalization of
the density matrix $\rho^{~}_{ab} \equiv {(C^4_{~})}_{ab}^{~}$. 
Let us consider the eigenvalue equation for the density matrix
\begin{equation}
\sum_{b} \rho^{~}_{ab} A^{\alpha}_{b} =
\lambda_{\alpha}^{~} A^{\alpha}_{a} \, ,
\end{equation}
where $\lambda_{\alpha}^{~}$ is the eigenvalue in decreasing order
$\lambda_{1} \geq \lambda_{2} \geq \ldots \geq 0$, and
${\bf A}^{\alpha}_{~} = ( A^{\alpha}_{1}, A^{\alpha}_{2}, \ldots )^T_{~}$ is
the corresponding eigenvector that satisfies the orthogonal  relation
\begin{equation}
\left( {\bf A}^{\alpha}_{~}, \, {\bf A}^{\beta}_{~} \right) =
\sum_{a} A^{\alpha}_{a} A^{\beta}_{a} = \delta^{\alpha \beta}_{~} \, .
\end{equation}
It has been known that $\lambda_{\alpha}$ rapidly approaches to zero
with respect to $\alpha$,~\cite{Bx3,Wh2} and that we can neglect tiny
eigenvalues from the view point of numerical calculation. We consider only
$m$ numbers of dominant eigenvalues in the following; the greek indices
run from $1$ to $m$. The number $m$ is determined so that
$\sum_{\alpha=1}^m \lambda_{\alpha}$ becomes a good lower bound for
the partition function $Z_{2N}^{~} = {\rm Tr} \, \rho$. 

Equation (2.4) shows that for a sufficiently large $m$ the
density matrix $\rho$ can be well approximated as
\begin{equation}
\rho^{~}_{ab} \sim
\sum_{\alpha=1}^m A^{\alpha}_{a} A^{\alpha}_{b} 
\lambda_{\alpha} \, .
\end{equation}
The above approximation shows that the $m$-dimensional diagonal matrix
\begin{equation}
{\tilde \rho}^{~}_{\alpha \beta} =
\sum^{~}_{a b} A^{\alpha}_{a} A^{\beta}_b {\rho}^{~}_{a b} = 
\delta_{\alpha \beta} \lambda_{\beta} 
\end{equation}
contains the relevant information of $\rho$; we can regard ${\tilde \rho}$
as the renormalized density matrix. This is the heart of the density
matrix algorithm: {\it the RG transformation is defined by
the matrix $A = \left( {\bf A}^{1}_{~}, {\bf A}^{2}_{~}, \ldots, {\bf
A}^{m}_{~} \right)$ which is obtained through the diagonalization of the
density matrix.}

As we have applied the RG transformation to the density matrix
$\rho$, we can renormalize the CTM by applying the matrix $A$ as 
\begin{equation}
{\tilde C}^{~}_{\alpha \beta} =
\sum^{~}_{a b} A^{\alpha}_{a} A^{\beta}_b C^{~}_{a b} \, .
\end{equation}
Since $C^{~}_{a b}$ and $\rho^{~}_{a b}$ have the common
eigenvectors --- remember that $\rho = C^4_{~}$ --- the renormalized
CTM is an $m$-dimensional diagonal matrix 
\begin{equation}
{\tilde C} = {\rm diag}(\omega_1, \omega_2, \ldots, \omega_m) \, ,
\end{equation}
where $\omega_{\alpha}$ is the eigenvalue of the CTM that satisfies
$\lambda_{\alpha} = \omega_{\alpha}^4$. 
In the same manner, we obtain the renormalized HRTM
\begin{equation}
{\tilde P}^{~i}_{\alpha \beta} = \sum^{~}_{a b}
A^{\alpha}_{a} A^{\beta}_{b} P^{~i}_{ab} \, .
\end{equation}
In this case ${\tilde P}^{~i}_{\alpha \beta}$ is not diagonal
with respect to $\alpha$ and $\beta$; {\it the RG transformation is not
always diagonalization.}

We can extend the linear size of CTM and HRTM using Eqs.(2.2) and (2.3),
and we can reduce their matrix dimension by the RG transformation in
Eqs.(2.7) and (2.8). By repeating the extension and the renormalization, we
can obtain the renormalized density matrix ${\tilde \rho}$ and the
approximate partition function ${\tilde Z}_{2N}^{~} = {\rm Tr} \, {\tilde
\rho}$ for arbitrary system size $N$. This is the outline of the CTMRG.

\section{Density matrix in Three Dimension}

In order to generalize the density matrix algorithm to 3D systems, we
first construct the density matrix in three dimension. As an example of 3D
systems, we consider a 64-vertex model, that is defined by a
Boltzmann weight $W^{~}_{ijklmn}$. (Fig.4(a)) In order to simplify the
following discussion, we consider the case where $W^{~}_{ijklmn}$ is
invariant under the permutations of the two-state spins $i, j, k, l, m$ and
$n$.~\cite{enough} As we have considered a square cluster in
two-dimension, (Fig.1) we consider a cube with linear dimension $2N$,
where the boundary spins (on the surface of the cube) are fixed to the
same direction. According to the variational formulation shown in \S 1,
we first decompose the cube into several parts shown in Fig.4(b)-(d). 

\begin{figure}
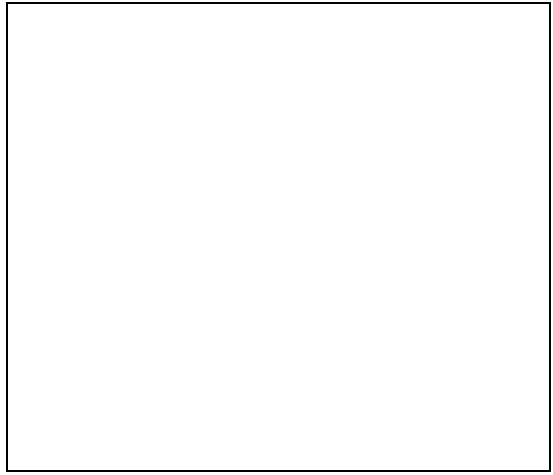

\figureheight{6cm}
\caption{
Parts of the cubic cluster with linear dimension $2N$:
(a) Vertex weight $W^{~}_{ijklmn}$. 
(b) The tensor $P^{~i}_{abcd}$.
(c) The tensor $S^{XY}_{ab}$.
(d) Corner Tensor $C_{~}^{XYZ}$.
The cross marks $\times$ represent the boundary spins. 
 } 
\label{fig:4}
\end{figure}

The tensor $P^{~i}_{abcd}$ shown in Fig.4(b) is a kind of
three-dimensional HRTM. The superscript $i$ represents the two-state
spin at the top. The spin at the bottom is fixed, because it is at the
boundary of the system. The subscript $a$ represents the group of in-line
spins $a = (a_1^{~}, a_2^{~}, \ldots, a_N^{~})$; $b$, $c$, and $d$ are
defined in the same way. From the symmetry of the vertex weight, 
$P^{~i}_{abcd}$ is invariant under the permutations of subscripts. 

The tensor $S^{XY}_{a \, b}$ shown in Fig.4(c) does not have its 2D
analogue; it is an array of vertices. 
The subscripts $a$ and $b$ represent in-line spins;
other two sides are the boundary of the cube. The
superscript $X$ represents an $N$ by $N$ array of spins
on the square surface
\begin{equation}
X = \left(
\begin{array}{cccc}
x_{11}^{~} & x_{12}^{~} & \ldots & x_{1N}^{~} \\
x_{21}^{~} & x_{22}^{~} & ~ & \vdots \\
\vdots & ~ & \ddots & \vdots \\
x_{N1}^{~} & \cdots & \cdots & x_{NN}^{~}
\end{array}
\right) \, ,
\end{equation}
where $x_{NN}^{~}$ is closest to the center of the cube, and $Y$ is the
spin array on the other surface; $x_{ij}^{~}$ and $y_{ij}^{~}$ are
connected to the same vertex at the position $\{ij\}$. The tensor is
invariant under the permutation of $X$ and $Y$ ($S^{XY}_{a \, b} = S^{YX}_{a
\, b}$), but is not invariant for $a$ and $b$ ($S^{XY}_{a \, b} \neq S^{XY}_{b
\, a}$); $S^{XY}_{a \, b}$ is equal to $S^{ZW}_{b \, a}$ where $Z = X^T_{~}$
and $W =  Y^T_{~}$.

Figure 4(d) shows the corner tensor $C^{XYZ}_{~}$, which is a kind
of three-dimensional CTM.~\cite{Bx3} The superscripts are defined in the
same way as Eq.(3.1). (The boundary spins on the surfaces of the original
cube are fixed.) It should be noted that $C^{XYZ}_{~}$ is not
equal to $C^{WYZ}_{~}$ where $W = X^T_{~}$, because each surface has
its own orientation.

\begin{figure}
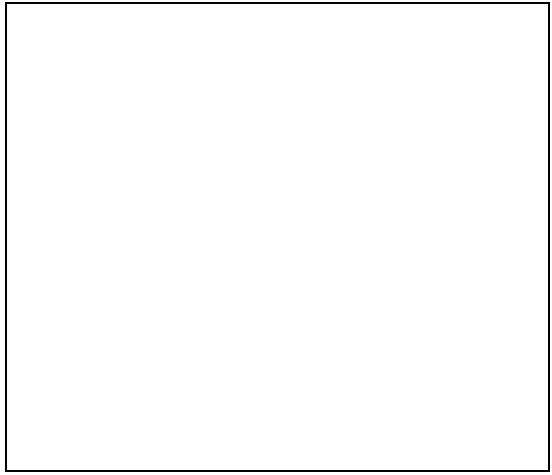

\figureheight{6cm}
\caption{
Extensions of (a) $P$ in Eq.(3.2), (b) $S$ in Eq.(3.3), and (c) $C$
in Eq.(3.5).
 } 
\label{fig:5}
\end{figure}

Following the formulation in two-dimension, let us consider the
size extension of $P$, $S$, and $C$. The length of $P$ can be
increased by joining a vertex (Fig.5(a))
\begin{equation}
P^{~i}_{{\bar a}{\bar b}{\bar c}{\bar d}} = 
\sum_n W^{~}_{ijklmn} \, P^{~n}_{abcd} \, ,
\end{equation}
where the extended in-line spins are defined as
${\bar a} = (a, j) = (a_1^{~}, a_2^{~}, \ldots, a_N^{~}, j)$, 
${\bar b} = (b, k) = (b_1^{~}, b_2^{~}, \ldots, b_N^{~}, k)$, 
${\bar c} = (c, l) = (c_1^{~}, c_2^{~}, \ldots, c_N^{~}, l)$, and
${\bar d} = (d, m) = (d_1^{~}, d_2^{~}, \ldots, d_N^{~}, m)$. 
The linear size of $S$ can be increased by joining two $P$
and a vertex (Fig.5(b))
\begin{equation}
S^{{\bar X}{\bar Y}}_{{\bar a} \, {\bar b}} = 
\sum_{ln} \sum_{ce} W^{~}_{ijklmn} \, 
P^{~n}_{abcd} \, P^{~l}_{efgh} \, S^{XY}_{c \, e}
\end{equation}
where the extended in-line spins are defined as 
${\bar a} = (a, j) = (a_1^{~}, a_2^{~}, \ldots, a_N^{~}, j)$, and 
${\bar b} = (g, i) = (g_1^{~}, g_2^{~}, \ldots, g_N^{~}, i)$.  
The extended spin array ${\bar X}$ is defined as 
\begin{equation}
{\bar X} = \left(
\begin{array}{ccc|c}
x_{11}^{~} & \ldots & x_{1N}^{~} & f_{1}^{~} \\
\vdots & \ddots & \vdots & \vdots \\
x_{N1}^{~} & \ldots & x_{NN}^{~} & f_{N}^{~} \\ \cline{1-4}
b_{1}^{~} & \ldots & b_{N}^{~} & k
\end{array}
\right) \, ,
\end{equation}
and ${\bar Y}$ is defined in the same way from the indices $m$, $d$, $h$
and $Y$. The linear size of the corner tensor $C$ can be increased by
joining three $P$, three $S$, and a vertex (Fig.5(c))
\begin{equation}
C^{{\bar X}{\bar Y}{\bar Z}}_{~} = 
\sum_{lmn} \sum_{cd \, eh \, qr} \sum_{TUV} 
W^{~}_{ijklmn} \, P^{~n}_{abcd} \, P^{~l}_{efgh} \, P^{~m}_{opqr} 
S^{XT}_{q \, d} \, S^{YU}_{c \, e} \, S^{ZV}_{h \, r} \, 
C^{TUV}_{~} \, .
\end{equation}
The extended superscripts ${\bar X}$,  ${\bar Y}$, and ${\bar Z}$ are
defined in the  same way as Eq.(3.4). In equation (3.5) we have to take
care of the orientation of the surfaces $T$, $U$, and $V$.

\begin{figure}
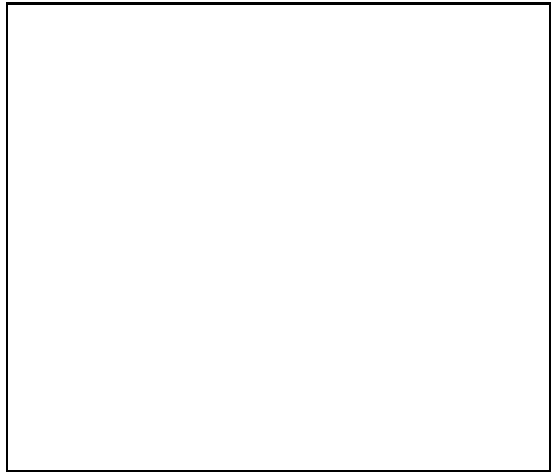

\figureheight{6cm}
\caption{
The density matrix $Q$ in Eq.(3.8) is obtained by joining two corner
tensors (Eq.(3.6)) to obtain the tensor $D$, and then joining
four of them.(Eq.(3.8))
 } 
\label{fig:6}
\end{figure}

Now we can express the partition function ${\Xi}_{2N}^{~}$ of the cube
with linear size $2N$ using the corner tensors. We first join two
corner tensors (Fig.5) to obtain a symmetric matrix
\begin{equation}
D^{(XU)(ZV)}_{~} = \sum_Y C^{XYZ}_{~} \, C^{UYV}_{\rm m} \, ,
\end{equation}
where we regard the pair $(ZV)$ as the column index of $D$, and $(XU)$ as
the row index. The tensor $C^{~}_{\rm m}$ is the mirror image of $C$:
$C^{UYV}_{\rm m} \equiv C^{{U^T_{~}}Y{V^T_{~}}}_{~}$.
The partition function ${\Xi}_{2N}^{~}$ is then expressed as 
\begin{equation}
{\Xi}_{2N}^{~} = {\rm Tr} \, D^4_{~} = \sum_{XU} Q^{(XU)(XU)}_{~} \, ,
\end{equation}
where the matrix $Q$ is the forth power of $D$ (Fig.6)
\begin{equation}
Q^{(XU)(ZV)}_{~} = \sum_{(AB)(CD)(EF)} 
D^{(XU)(AB)}_{~} \, D^{(AB)(CD)}_{~} \, 
D^{(CD)(EF)}_{~} \, D^{(EF)(ZV)}_{~} \, .
\end{equation}
The matrix $Q$ can be seen as a density matrix for the cube, 
because ${\rm Tr} \, Q$ is the partition function
${\Xi}_{2N}^{~}$. By contracting two superscripts of $Q$, we obtain a
density submatrix
\begin{equation}
\rho^{XZ}_{~} = \sum_U Q^{(XU)(ZU)}_{~} \, ,
\end{equation}
which will be used for the RG transformation for the spin array.

Let us consider a density submatrix $\rho^{{\bar X}{\bar Z}}_{~}$ for the 
extended cube with size $2(N + 1)$, where ${\bar X}$ is the extended
spin array Eq.(3.4); for a while we label the elements of ${\bar
Z}$ as

\begin{equation}
{\bar Z} = \left(
\begin{array}{ccc|c}
{x'}_{11}^{~} & \ldots & {x'}_{1N}^{~} & {f'}_{1}^{~} \\
\vdots & \ddots & \vdots & \vdots \\
{x'}_{N1}^{~} & \ldots & {x'}_{NN}^{~} & {f'}_{N}^{~} \\ \cline{1-4}
{b'}_{1}^{~} & \ldots & {b'}_{N}^{~} & {k'}
\end{array}
\right) \, 
\end{equation}
in order to define another density submatrix.
By tracing out $N$ by $N + 1$ variables of the extended density matrix
$\rho^{{\bar X}{\bar Z}}_{~}$
\begin{equation}
\rho_{{\bar f}{\bar g}}^{~} 
= \sum_{b_i^{~} = {b'}_i^{~}~~x_{ij}^{~} = {x'}_{ij}^{~}}
\rho^{{\bar X}{\bar Z}}_{~} \, ,
\end{equation}
where 
${\bar f} = (f_1^{~}, \ldots, f_N^{~}, k)$ and 
${\bar g} = ({f'}_1^{~}, \ldots, {f'}_N^{~}, k')$, we obtain another density
submatrix for the extended in-line spins. In the same way, we obtain
$\rho_{fg}^{~}$ for the in-line spins of length $N$ --- 
$f = (x_{1N}^{~}, \ldots, x_{NN}^{~})$ and  
$g = ({x'}_{1N}^{~}, \ldots, {x'}_{NN}^{~})$ --- 
by tracing out $N-1$ by $N$ variables of $\rho^{XZ}_{~}$ in Eq.(3.9).

\section{RG Algorithm in Three Dimension}

As we have done in \S 2, we  obtain RG transformations by way of the
diagonalizations of the density submatrices. We first consider the
eigenvalue relation 
\begin{equation}
\sum_Z^{~} \rho^{XZ}_{~} U^Z_{\Psi} = \Lambda_{\Psi} U^X_{\Psi} \, ,
\end{equation}
where we assume the decreasing order for $\Lambda_{\Psi}$. We keep
first $m'$ eigenvalues, ($\Psi = 1, \ldots, m'$) and neglect the rest of
relatively small ones. We then obtain the RG transformation matrix
$U^X_{\Psi}$, that maps the spin array $X$ to an $m'$-state block spin
$\Psi$. For example, the corner tensor $C^{XYZ}_{~}$ is renormalized as
(Fig.7(a))
\begin{equation}
{\tilde C}^{\Psi \Phi \Theta}_{~} = \sum_{XYZ}^{~}
U_{\Psi}^X U_{\Phi}^Y U_{\Theta}^Z C^{XYZ}_{~} \, .
\end{equation}
It should be noted that under the transpose of the spin array $X
\rightarrow X^T_{~}$ the matrix $U^X_{\Psi}$ transforms as 
$\pm U^X_{\Psi}$ according to the parity of the block spin $\Psi$.

Let us consider another eigenvalue relation 
\begin{equation}
\sum_g^{~} \rho_{fg}^{~} A_g^{\psi} = 
\lambda_{~}^{\psi} A_f^{\psi}
\end{equation}
for the density submatrix $\rho_{fg}^{~}$, where $f$ and $g$ are in-line
spins. We assume
the decreasing order for $\lambda^{\psi}_{~}$ as before, and we keep $m$
numbers of large eigenvalues. ($\psi = 1, \ldots, m$) The matrix
$A_f^{\psi}$ then represent the RG transformation for the in-line spin $f$.
For example, $P^{~i}_{abcd}$ is renormalized as (Fig.7(b))
\begin{equation}
{\tilde P}^{~i}_{\alpha \beta \gamma \delta} = \sum_{abcd}^{~}
P^{~i}_{abcd} A^{\alpha}_a A^{\beta}_b A^{\gamma}_c A^{\delta}_d \, .
\end{equation}
By using both $U^X_{\Psi}$ and $A_f^{\psi}$, we can renormalize 
$S_{ab}^{XY}$ as (Fig.7(c))
\begin{equation}
{\tilde S}_{\alpha \beta}^{\Psi \Phi} = \sum_{a b X Y}
S_{ab}^{XY} A_a^{\alpha} A_b^{\beta} U_{\Psi}^X U_{\Phi}^Y \, .
\end{equation}
As a result of RG transformations, the tensors $P^{~i}_{abcd}$,
$S_{ab}^{XY}$, and $C_{~}^{XYZ}$ are approximated as
\begin{eqnarray}
P^{~i}_{abcd} & \sim & \sum_{\alpha \beta \gamma \delta = 1}^m
 A^{\alpha}_{a} A^{\beta}_{b} A^{\gamma}_{c} A^{\delta}_{d}
{\tilde P}^{~i}_{\alpha \beta \gamma \delta} \\
S^{XY}_{a \, b} & \sim & \sum_{\alpha \beta = 1}^m 
\sum_{\Psi \Phi =1}^{m'}
 A^{\alpha}_{a} A^{\beta}_{b} U^{X}_{\Psi} U^{Y}_{\Phi}
{\tilde S}^{\Psi \Phi}_{\alpha \, \beta} \\
C^{XYZ}_{~} & \sim & \sum_{\Psi \Phi \Theta = 1}^{m'}
 U^{X}_{\Psi} U^{Y}_{\Phi} U^{Z}_{\Theta}
{\tilde C}^{\Psi \Phi \Theta}_{~} \, .
\end{eqnarray}
For the models that have unique ground-state spin configuration,
the above approximations become exact when $T = 0$ and $T = \infty$ even
for $m = m' = 1$.

\begin{figure}
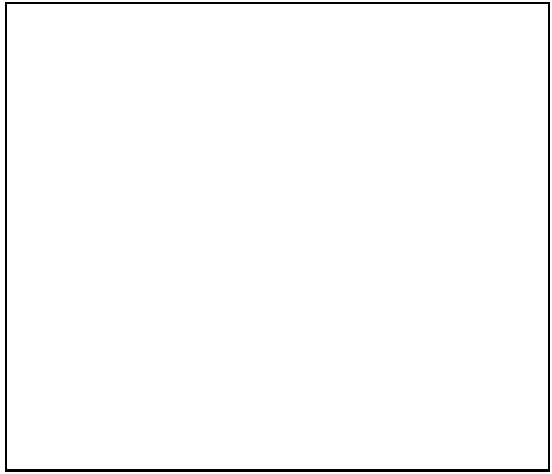

\figureheight{6cm}
\caption{
The renormalized tensors 
(a) ${\tilde P}^{~i}_{\alpha \beta \gamma \delta}$ in Eq.(4.4),
(b) ${\tilde S}_{\alpha \beta}^{\Psi \Phi}$ in Eq.(4.5), and 
(c) ${\tilde C}^{\Psi \Phi \Theta}_{~}$ in Eq.(4.2).
The greek letters $\alpha$, $\beta$, $\gamma$, and $\delta$ denote
$m$-state renormalized in-line spins, and the capital ones $\Psi$, $\Phi$,
and $\Theta$ denote $m'$-state renormalized spin arrays.
 } 
\label{fig:7}
\end{figure}

Now we can directly generalize the algorithm of CTMRG to 3D lattice
models. The algorithm consists of the extensions for
$P^{~i}_{abcd}$ (Eq.(3.2)), $S^{XY}_{ab}$ (Eq.3.3)), and 
$C^{XYZ}_{~}$ (Eq.(3.5)), and the RG transformations Eqs.(4.2),(4.4) and
(4.5).  The procedure of the renormalization group is as follows:
\begin{itemize}
\item[(1)] Start from $N = 1$, where all the tensors can be expressed by
the Boltzmann weight $W_{ijklmn}^{~}$: 
$P^{~i}_{abcd} = W_{iabcd \times}^{~}$, 
$S^{XY}_{ab} = W_{aXbY \times \times}^{~}$, and
$C^{XYZ}_{~} = W_{ZXY \times \times \times}^{~}$, where the mark
`$\times$' represents the fixed boundary spin.
\item[(2)] Join the tensors $W_{ijklmn}^{~}$, $P^{~i}_{abcd}$, 
$S^{XY}_{ab}$, and $C^{XYZ}_{~}$ using Eqs.(3.2), (3.3), and (3.5),
respectively, and obtain the extended ones $P^{~i}_{{\bar a}{\bar b}{\bar
c}{\bar d}}$, $S^{{\bar X}{\bar Y}}_{{\bar a}{\bar b}}$, and $C^{{\bar X}{\bar
Y}{\bar Z}}_{~}$. (Increment $N$ by one.)
\item[(3)] Using the extended corner tensor $C^{{\bar X}{\bar Y}{\bar
Z}}_{~}$ in Eq.(3.5), calculate the density matrix $\rho^{{\bar X}{\bar
Z}}_{~}$ via Eq.(3.9) and its submatrix $\rho^{~}_{{\bar f}{\bar g}}$ in
Eq.(3.11). 
\item[(4)] Obtain the RG transformation matrix $U_{\Psi}^{\bar X}$ and 
$A_{\bar a}^{\alpha}$ using Eqs.(4.1) and (4.3), respectively. We keep 
$m'$ states for $\Psi$, and $m$ states for $\alpha$.
\item[(5)] Apply the RG transformations to the extended tensors to
obtain ${\tilde P}^{~i}_{\alpha \beta \gamma \delta}$ (Eq.(4.4)),
${\tilde S}_{\alpha \beta}^{\Psi \Phi}$ (Eq.(4.5)), and
${\tilde C}^{\Psi \Phi \Theta}_{~}$ (Eq.(4.2)).
\item[(6)] Goto the step (2) and repeat the procedures (2)-(5)
for the renormalized tensors ${\tilde P}$, ${\tilde S}$, and ${\tilde C}$.
\end{itemize}
Every time we extend the tensors in the step (2) the system size ---
the linear dimension of the cube --- increases by 2. After the step (3) we
can obtain the lower bound of the partition function by taking the trace of
the density submatrix $\Xi_{2(N+1)}^{~} = {\rm Tr} \, \rho^{{\bar X}{\bar
Z}}_{~} = {\rm Tr} \, \rho^{~}_{{\bar f}{\bar g}}$. We stop the iteration
when the partition function per vertex converges. Since the extended spin
array ${\bar X}$ of the density matrix $\rho^{{\bar X}{\bar Z}}_{~}$
contains the original (unrenormalized) spin variable, we can directly
calculate the local energy and the order parameter.~\cite{CTMRG2}

Let us apply the above algorithm to the 3D Ising model. The model
is equivalent to the 64-vertex model whose vertex weight is
given by 
\begin{equation}
W_{ijklmn}^{~} = \sum_{\sigma = \pm 1}^{~}
U_{\sigma i}^{~} U_{\sigma j}^{~} U_{\sigma k}^{~}
U_{\sigma l}^{~}U_{\sigma m}^{~}U_{\sigma n}^{~} \, ,
\end{equation}
where $U_{\sigma i}^{~}$ is unity when $\sigma = i$, and is
$e^K_{~} + \sqrt{e^{2K}_{~}-1}$ when $\sigma \ne i$. The parameter $K$
denotes the inverse temperature $J / k_{\rm B} T$. 
For this model the initial conditions for step (1) are slightly modified as
\begin{eqnarray}
P^{~i}_{abcd} & = &  U_{\times i}^{~}
U_{\times a}^{~} U_{\times b}^{~}
U_{\times c}^{~} U_{\times d}^{~}  \nonumber\\
S^{XY}_{ab} & =  & 
U_{\times X}^{~} U_{\times Y}^{~} 
U_{\times a}^{~} U_{\times b}^{~}  \nonumber\\
C^{XYZ}_{~} & = & U_{\times X}^{~}  U_{\times Y}^{~}  U_{\times Z}^{~}  \, ,
\end{eqnarray}
where `$\times$' represent the boundary Ising spin. (The modification is
nothing but the change in normalizations.) We impose ferromagnetic
boundary condition $\times = 1$. As a trial calculation we keep only two
states for both in-line spins ($m = 2$) and spin arrays ($m' = 2$); when
$m' = 2$ the parity of the renormalized spin array $\Psi$ in Eq.(4.1) is
always even. Figure 8 shows the calculated spontaneous magnetization.
Because of the smallness of $m$ and $m'$, the transition temperature
$T_{\rm c}^{~}$ is over estimated, where the feature is
common to the Klamers-Wannier approximation.~\cite{Krm}

\begin{figure}
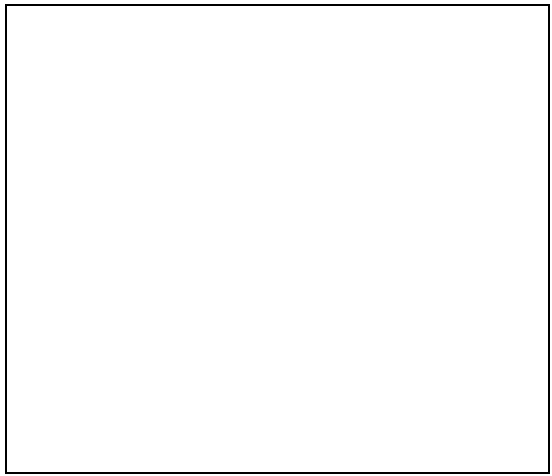

\figureheight{6cm}
\caption{
Calculated spontaneous magnetization of the 3D Ising model when $m =
m' = 2$. The arrow shows the true $T_{\rm c}^{~}$.
 } 
\label{fig:8}
\end{figure}

Compare to the CTMRG algorithm for 2D classical systems, the above RG
algorithm for 3D system requires much more computational time. The
reason is that after the step (2) the extended in-line spin ${\bar f}$
becomes $2 m$-state, and the extended spin array ${\bar X}$ becomes $2
m^2 m'$-state; in order to obtain $\rho^{{\bar X}{\bar Z}}_{~}$ in the step
(3) we have to create a matrix $D^{({\bar X}{\bar U})({\bar Z}{\bar
V})}_{~}$ by Eq.(3.6), whose dimension is $4 m^4 {m'}^2$. For the simplest
(non-trivial) case $m = m' = 2$ the dimension is already 256. 

\section{Conclusion and discussion}

We have explained a way of generalizing the RG algorithm of
CTMRG~\cite{CTMRG1,CTMRG2} to 3D classical models, focusing on the
construction of the density matrix from eight corner tensors. The RG
transformations are obtained through the diagonalizations of the density
matrices. As far as we know it is the first generalization of the
infinite-system density matrix algorithm~\cite{Wh1,Wh2} to 3D classical
systems.

From the computational view point, the calculation in 3D is far more
heavy than that of CTMRG in 2D; we have to improve the numerical
algorithm in 3D for realistic use. What we have done is to approximate the
eigenstate of a transfer matrix in 3D as a two-dimensional product of
renormalized tensor ${\tilde P}$ (Eq.(4.4)); the most important process is
to improve the tensor elements in ${\tilde P}$ so that the variational
partition function is maximized. The improvement of tensor product
state for 1D quantum system proposed by Dukelsky et.
al.~\cite{RecGerman1,RecGerman2}, where their algorithm does not
explicitly require the density matrix, may of use to reduce the numerical
effort in three dimension.

The authors would like to express their sincere thanks to Y.~Akutsu,
M.~Kikuchi, for valuable discussions. T.~N. is greatful to G.~Sierra about
the discussion on the matrix product state. K.~O. is supported by JSPS
Research Fellowships for Young Scientists. The present work is partially
supported by a Grant-in-Aid from Ministry of Education, Science and Culture
of Japan. Most of the numerical calculations were done by NEC SX-4 in
computer center of Osaka University.


\begin{thebibliography}{99}
\bibitem{Krm} H.~A.~Kramers and G.~H.~Wannier, Phys. Rev. 
{\bf 60} (1941) 263.
\bibitem{Kik} R.~Kikuchi, Phys. Rev. {\bf 81} (1951) 988.
\bibitem{Bethe} H.~A.~Bethe: Proc. Roy. Soc. {\bf A150}
(1935) 552.
\bibitem{Bx1} R.~J.~Baxter: J. Math. Phys. {\bf 9} (1968) 650 .
\bibitem{Bx2} R.~J.~Baxter: J. Stat. Phys. {\bf 19} (1978) 461.
\bibitem{Bx3} R.~J.~Baxter: {\it Exactly Solved Models in
Statistical Mechanics} (Academic Press, London, 1982) p.363.
\bibitem{NB} N.~P.~Nightingale and H.~W.~Bl\"ote: Phys. Rev. 
{\bf B33} 659.
\bibitem{AKLT} I.~Affleck, T.~Kennedy, E.~H.~Lieb, and 
H.~Tasaki: Phys. Rev. Lett. {\bf 59} (1987) 799.
\bibitem{Fannes1} M.~Fannes, B.~Nachtergale, and R.~F.~Werner:
Europhys. Lett. {\bf 10} (1989) 633.
\bibitem{Fannes2} M.~Fannes, B.~Nachtergale, and R.~F.~Werner:
Commun. Math. Phys. {\bf 144} (1992) 443.
\bibitem{Fannes3} M.~Fannes, B.~Nachtergale, and R.~F.~Werner:
Commun. Math. Phys. {\bf 174} (1995) 477.
\bibitem{Zitt1} A.~Kl\"umper, A.~Schadschneider and J.~Zittarz:
Z. Phys. {\bf B87} (1992) 281.
\bibitem{Zitt2} H.~Niggemann, A.~Kl\"umper, and J.~Zittartz:
Z. Phys. {\bf B104} (1997) 103-110.
\bibitem{Takas} H.~Takasaki and T.~Nishino: unpublished.
\bibitem{Wh1} S.~R.~White: Phys. Rev. Lett. {\bf 69} (1992) 2863.
\bibitem{Wh2} S.~R.~White: Phys. Rev. {\bf B48} (1993) 10345.
\bibitem{Ostlund1} S.~\"Ostlund and S.~Rommer: Phys. Rev. Lett
{\bf 75} (1995) 3537.
\bibitem{Ostlund2} S.~Rommer and S.~\"Ostlund: Phys. Rev. {\bf B55} 
(1997) 2164.
\bibitem{Huse} S.~R.~White and D.~A.~Huse: Phys. Rev. 
{\bf B48} (1993) 3844.
\bibitem{Hida} K.~Hida: J. Phys. Soc. Jpn. {\bf 65} (1996) 895.
\bibitem{Escorial} G.~Sierra and M.~A.~Mart\'{\i}n-Delgado: 
{\it Strongly Correlated  Magnetic and Superconducting Systems,} 
(Springer Berlin, 1997), and references there in.
\bibitem{Ni} T.~Nishino, J. Phys. Soc. Jpn. {\bf 64}, No.10 (1995) 3598.
\bibitem{Carlon1} Enrico~Carlon and Andrej Drzewi\'nski: Phys. 
Rev. Lett. {\bf 79} (1997) 1591.
\bibitem{Carlon2} Enrico~Carlon and Andrej Drzewi\'nski:
cond-mat/9709176. 
\bibitem{RecGerman0} G.~Sierra, M.~A.~Mart\'{\i}n-Delgado,
J.~Dukelsky, S.~R.~White, and D.~J.~Scalapino:
cond-mat/9707335.
\bibitem{RecGerman1} J.~Dukelsky, M.~A.~Mart\'{\i}n-Delgado,
T.~Nishino, and G.~Sierra: cond-mat/9710310.
\bibitem{RecGerman2} J.~M.~Roman, G.~Sierra, J.~Dukelski,
and M.~A.~Mart\'{\i}n-Delgado: cond-mat/9802150.
\bibitem{CTMRG1} T.~Nishino and K.~Okunishi: J. Phys. Soc. Jpn.
{\bf 65} (1996) 891.
\bibitem{CTMRG2} T.~Nishino and K.~Okunishi: J. Phys. Soc. Jpn.
{\bf 66} (1997) 3040.
\bibitem{q5} T.~Nishino and K.~Okunishi: to be published in 
J. Phys. Soc. Jpn. {\bf 67} (1998); cond-mat/9711214.
\bibitem{Simple} The generalization of CTMRG to asymmetric vertex 
models is straight forward.
\bibitem{enough} To introduce such a symmetry is not a over
simplification. For example, the $d = 3$ Ising model can be 
expressed as a three-dimensional $64$-vertex model. 
(See Eqs.(4.9)-(4.12))
\end{thebibliography}
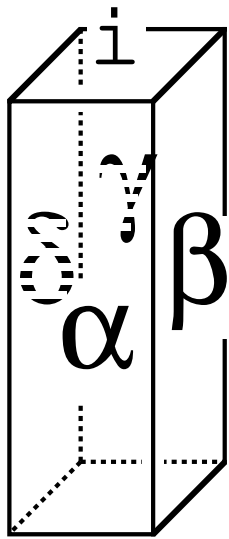
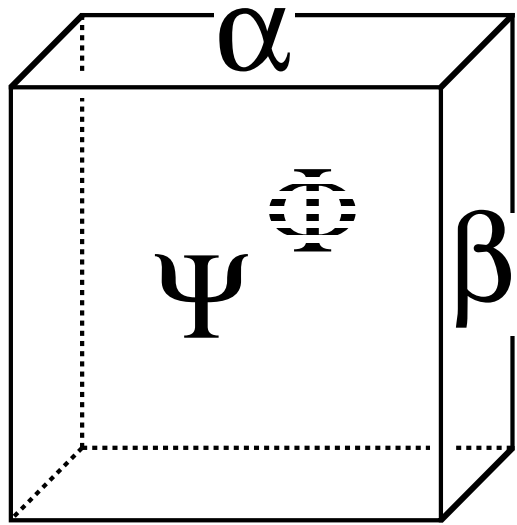
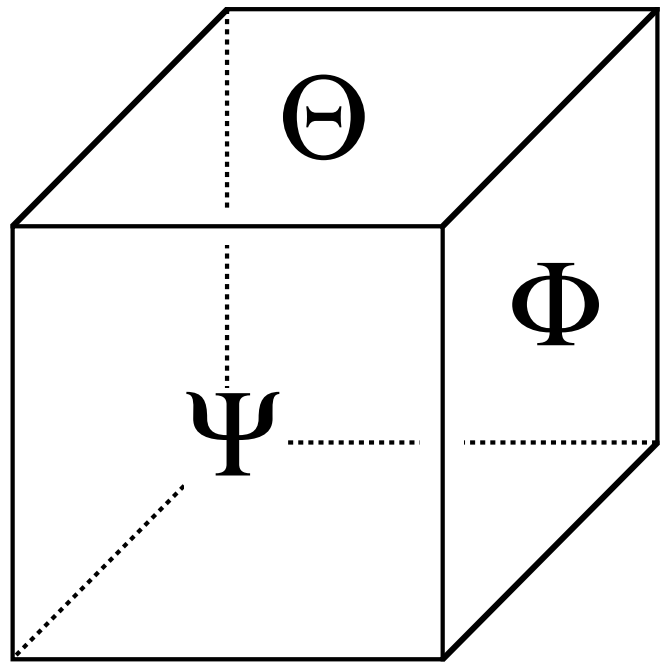
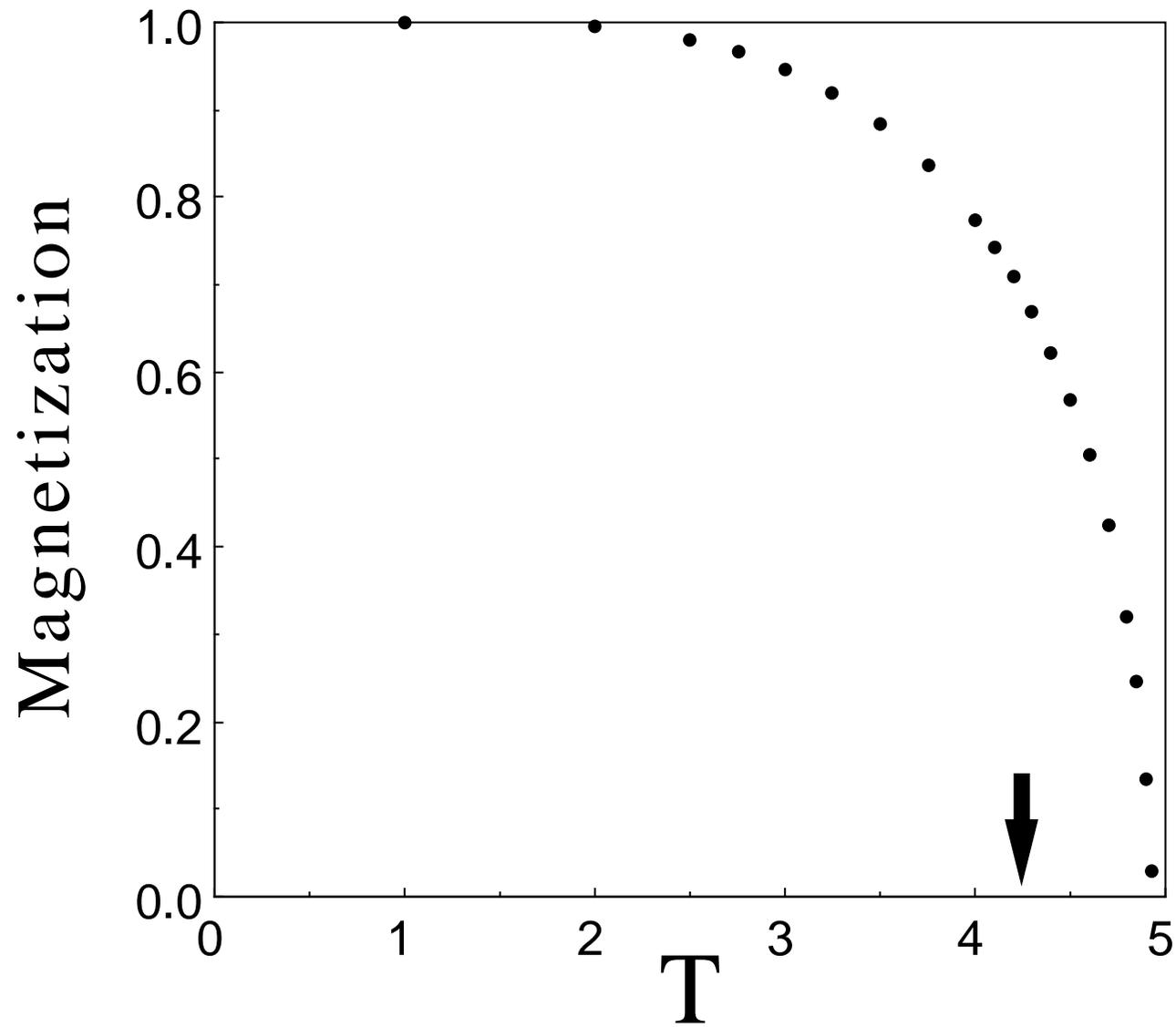
\end{document}